# Photonic Crystal Cavities in Cubic (3C) Polytype Silicon Carbide Films

Marina Radulaski,[1,*] Thomas M. Babinec,[1] Sonia Buckley,[1] Armand Rundquist,[1] J Provine,[2] Kassem Alassaad,[3] Gabriel Ferro,[3] and Jelena Vučković[1]

[1]*E. L. Ginzton Laboratory, Stanford University, Stanford, CA 94305, U.S.A.*
[2]*Department of Electrical Engineering, Stanford University, Stanford, CA 94305, U.S.A.*
[3]*Laboratorie des Multimateriaux et Interfaces, Universite de Lyon, 69622 Villeurbanne Cedex, France*
[*]*marina.radulaski@stanford.edu*

**Abstract:** We present the design, fabrication, and characterization of high quality factor ($Q \sim 10^3$) and small mode volume ($V \sim 0.75\,(\lambda/n)^3$) planar photonic crystal cavities from cubic (3C) thin films (thickness ~ 200 nm) of silicon carbide (SiC) grown epitaxially on a silicon substrate. We demonstrate cavity resonances across the telecommunications band, with wavelengths from 1.25 – 1.6 μm. Finally, we discuss possible applications in nonlinear optics, optical interconnects, and quantum information science.

## References and links

## 1. Introduction to Cubic (3C) Silicon Carbide Photonics

Wide band-gap semiconductors have recently emerged as an important material platform for nanophotonics and quantum information science, with applications including room temperature single photon sources and quantum nodes.[1-3] Among these wide band-gap materials, diamond has been studied most extensively in recent years because of the combination of its wide optical transparency window,[4,5] well-known and optically-addressable spin qubits,[6,7] and advanced materials processing techniques such as isotopic engineering.[8,9] While great advances in generating nanophotonic structures in single crystal diamond have been made,[10-15] certain material properties present some long-term challenge to integration and to application in classical and quantum photonic systems.

Silicon carbide (SiC), on the other hand, has recently been recognized as a material system with an advantageous combination of electronic, mechanical, and optical properties for these applications.[16] For example, SiC is non-centrosymmetric and differs from diamond and many other centrosymmetric materials with a large transparency window in the visible ($SiO_2$, $SiN_x$, etc.). Neglecting surface effects, only third order susceptibility ($\chi^{(3)}$) effects can be observed in these materials and this precludes the utilization of processes such as second harmonic generation, three wave mixing and the linear electro-optic (Pockels) effect. Conversely, semiconductors such as GaP[17-19], GaN,[20] and AlN[21], and crystalline materials such as $LiNbO_4$[22] have high second order susceptibilities. While microcavities in these materials have been used to perform low input power frequency conversion between near infrared and visible wavelengths with minimal absorption, many of these materials are not suitable for implementation of quantum nodes due to a lack of suitable quantum emitters, as well as inability to perform isotopic engineering for ultralong quantum memory times.

However, special considerations must be made in the development of silicon carbide photonics. This results from the fact that silicon carbide exhibits great material diversity and may exist in different crystal structures with different stacking sequences (polytypes). Optical devices based on these different SiC polytypes possess different materials processing and growth techniques, material properties, and spectrum of future applications. Crystal polytype is therefore a degree of freedom to be considered and taken advantage of in silicon carbide photonics. Specifically, we present in this paper the design, fabrication, and characterization of photonic crystal cavities in cubic (3C) SiC material. This 3C-SiC platform offers several practical advantages and fundamental opportunities.

One of the advantages of 3C-SiC lies in available materials growth techniques. For example, previous work on SiC photonic crystals employed bulk 6H-SiC material.[23,24] An application of the Smart-Cut technique[25] was therefore required to generate a device layer before photonic crystal patterning. Our approach here is to utilize 3C-SiC films grown directly on a Si substrate,[26] which allows for direct patterning in the 3C-SiC device layer without the requirement of ion implantation and damage. Such heteroepitaxial growth on sacrificial substrates, especially on silicon, has been recognized in other materials as an important step towards photonic integration.[27]

Moreover, the 3C-SiC employed in this work has very different optical properties relative to 6H-SiC used in the prior demonstration of SiC photonic crystals.[23,24] Due to crystal symmetry (e.g. cubic 3C vs. hexagonal 6H) nonlinear optical properties differ between the polytypes both in magnitude and in symmetry[28]. The symmetry of the nonlinear susceptibility tensors impacts applications in cavity-enhanced nonlinear optics. As an example, the 3C polytype grown in the [111] direction may allow for coupling between two TE modes, similar to GaAs which has the same cubic symmetry, in three wave mixing processes.[29] The different SiC polytypes are also known to contain distinct optically active crystalline defects (analogous to the NV center in diamond) for use in quantum information processing.[30]

Lastly, the photonic crystal platform presented in this paper differs from recent demonstrations of whispering gallery mode resonators in 3C-SiC.[31,32] Compared to the micron-scale ring resonator devices, the photonic crystal cavities demonstrated in this work may possess simultaneous high quality factors and small mode volume, which is necessary for applications such as cavity QED and quantum information processing.[33]

The paper is organized as follows: in section 2 we demonstrate that our materials and fabrication method is versatile and easily applied to diverse photonic crystal devices such as photonic crystal waveguides and cavities for integration with embedded quantum emitters,[34] 1D nanobeam structures for opto-mechanics,[35] and crossbeam cavities for nonlinear optics.[18] In section 3 we provide a demonstration of high quality factor ($Q \sim 10^3$) resonances in 2D photonic crystal cavities at telecommunications wavelengths from 1,250 – 1,600 nm.

## 2. Fabrication of Diverse Photonic Crystal Cavities in 3C-SiC Films

Fig. 1 shows the fabrication process flow we developed to realize a diverse set of 3C-SiC photonic crystal cavities. The starting material consists of a 3C-SiC layer grown epitaxially on a 35 mm diameter Si substrate using a standard two-step chemical vapor deposition (CVD) process. The film thickness $d \sim 210$ nm and refractive index ($n \sim 2.55$ at 1,300 – 1,500 nm) were confirmed using a Woollam spectroscopic ellipsometer before proceeding to subsequent masking and etching steps. A $SiO_2$ hard mask was then deposited using a low-pressure CVD furnace followed by $SiO_2$ thinning in a 2% HF solution to the target thickness of ~ 200 nm. Next, a second mask layer (Fig. 1a) of ~ 400 nm of electron beam lithography resist (ZEP 520a) was spun. Photonic crystal cavities were then patterned at the base-dose range 250 – 400 μC/cm$^2$ using a 100 keV electron beam lithography tool (JEOL JBX 6300) and then developed using standard recipes (Fig. 1b-c). The initial pattern transfer from resist into $SiO_2$ (Fig. 1d) was performed using $CF_4$, $CHF_3$, and Ar chemistry in a plasma etching system (Advanced Materials, p5000). Etch tests on bare (un-patterned) chips indicated that the selectivity of this etch was approximately 1:1 (SiO2 to ZEP). Residual electron beam resist was then completely removed in Microposit remover 1165 (Fig. 1e), and a second pattern transfer from SiO2 into SiC using HBr and Cl2 chemistry (Fig. 1f) was performed in the same system. Etch tests on bare chips indicated that the selectivity was approximately 2:1 (SiC to SiO$_2$). Next, undercutting of the SiC device layer was achieved by isotropic etching of the underlying Si (Fig. 1g) with a $XeF_2$ vapor phase etcher (Xactix e-1). Finally, the remaining $SiO_2$ hard mask layer was removed in 2% HF solution (Fig. 1h).

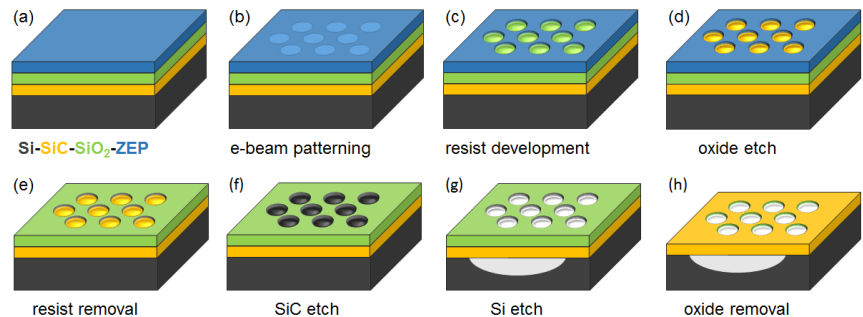

Fig. 1. The process flow to generate suspended photonic crystal structures in 3C-SiC films.

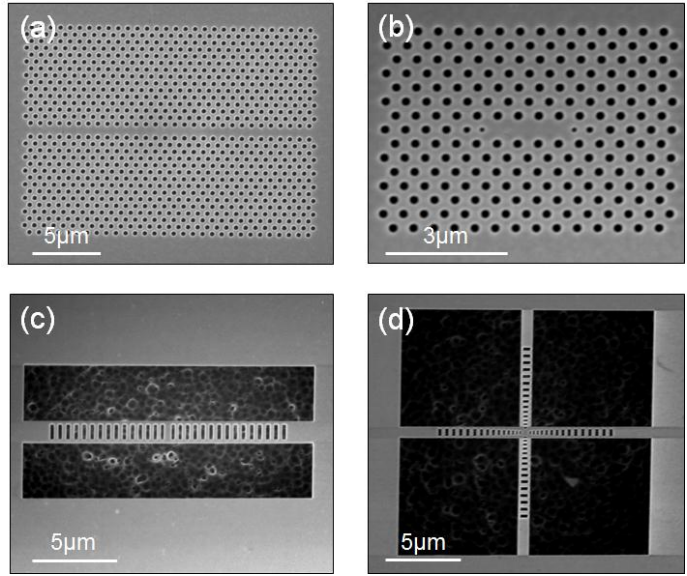

Fig. 2. Photonic crystal cavities fabricated from 3C-SiC films on a silicon substrate using the process flow in Fig. 1.
(a) Photonic crystal waveguides. (b) Planar L3 photonic crystal cavity. (c) 1D nanobeam photonic crystal cavity.
(d) Crossbeam photonic crystal cavity.

This fabrication procedure is suitable for various photonic crystal structures. Several different geometries are presented in Fig. 2. For example, Fig. 2a shows a realization of a 2D photonic crystal waveguide and 2b a planar photonic crystal cavity, which may be integrated with color centers in the future, similarly to the development of quantum photonic networks based on InAs quantum dots in GaAs.[34] Fig. 2c shows an example of a tapered 1D nanobeam cavity for applications in opto-mechanics.[35] Finally, Fig. 2d shows an example of a crossbeam cavity which may, in principle, support orthogonal modes with high spatial overlap for frequency conversion in nonlinear optics.[18]

## 3. Planar SiC Photonic Crystal Cavities at Telecommunications Wavelengths

In order to demonstrate properties of 3C-SiC platform for photonics applications, we proceed to design, fabricate and characterize low mode-volume, high quality factor, resonances at telecommunications wavelengths. In particular, the devices examined in this paper are planar photonic crystal cavities implementing a shifted linear three-hole defect (L3). These cavities (Fig. 3) have their three central holes removed and holes adjacent the cavity center displaced outwards by 15% of the lattice constant. The structures are also characterized by a device layer thickness $d \sim 210$ nm, silicon carbide refractive index $n \sim 2.55$, and hole radius $r = 0.3a$. Finally, for our initial purposes of prototyping devices in the 3C-SiC films, we choose a simple and robust fabrication design where all photonic crystal holes are the same size. Finite-difference time-domain (FDTD) simulations indicate that these structures possess theoretical mode volume values $V \sim 0.75 \, (\lambda/n)^3$ and cavity quality factors $Q \sim 2,000$.

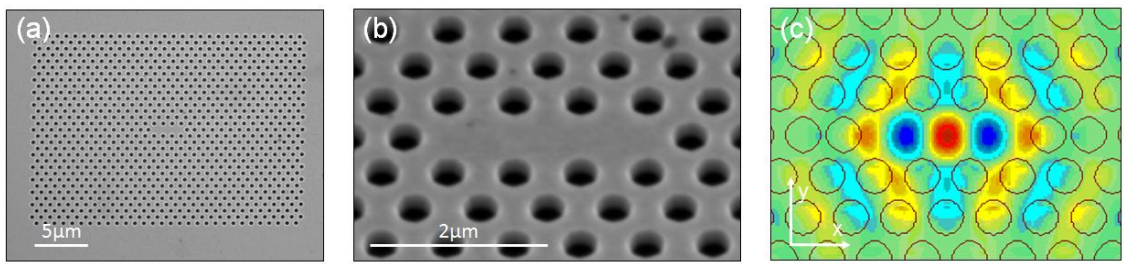

Fig. 3. SEM of fabricated 3C-SiC shifted L3 cavity from (a) orthogonal incidence, and (b) 40-degree tilt. (c) FDTD simulation of the dominant ($E_y$) electric field component of the fundamental mode, with the outlines of the holes indicated.

The fabricated 3C-SiC photonic crystal cavities were characterized using a cross-polarization measurement technique with apparatus presented in Fig. 4a. In this experiment, a halogen lamp (Ocean Optics) provided broadband excitation of the L3 cavity resonances in the wavelength region 1,250 – 1,600 nm and a high numerical aperture (*NA* = 0.5) objective lens was used to optically address the cavity with high efficiency. Incoming light was first polarized vertically (*y*-direction in Fig 3c) and reflected off the polarizing beam splitter (PBS) towards the sample of L3 cavities oriented horizontally and with cavity resonance polarized vertically. A half wave plate (HWP), with optical axis oriented at $\theta$ with respect to the vertical, was positioned between the PBS and objective. Resonance spectra were monitored via transmission through the PBS on a liquid nitrogen-cooled InGaAs spectrometer (Princeton Instruments).

Rotating the HWP modifies the cavity coupled light, as well as transmission through the PBS, in order to verify the polarization properties of the L3 photonic crystal cavity. For example, even though the cavity coupling is maximal for a half wave plate oriented at $\theta = 0°$, signal transmitted through the PBS is still zero. At $\theta = 45°$, both cavity coupling as well as signal transmitted through the PBS is zero. A high signal/noise ratio of cavity coupled light to uncoupled background is achieved when the HWP at $\theta = 22.5°$ rotates the polarization of the incoming beam at 45° from the vertical. Fig. 4b shows one such spectrum of the fundamental mode of a representative 3C-SiC cavity at 1,530 nm with quality factor of 800, obtained from a fit to Fano-resonance.[36] The peak intensity of the cavity resonance has $\sin(4\theta)$ dependence and the polarization-dependent signal presented in Fig. 4c shows good agreement with this theory.

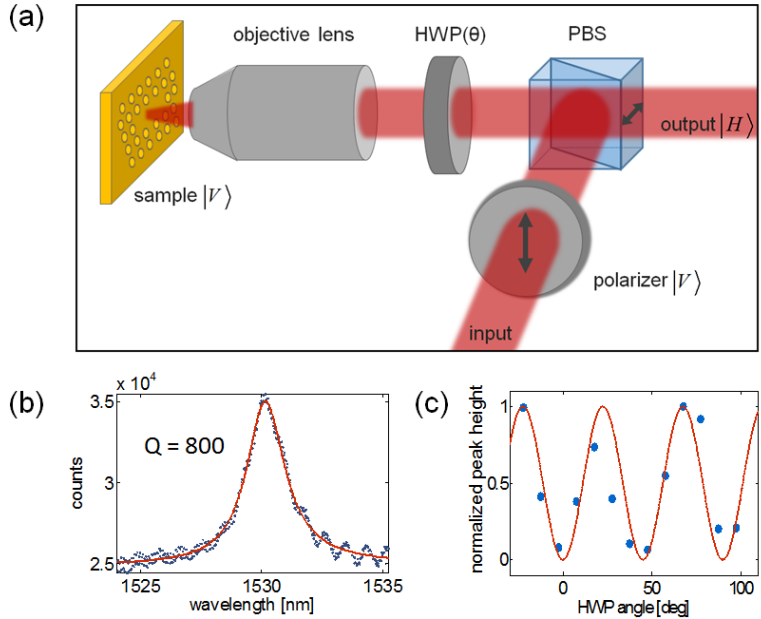

Fig. 4. (a) Cross-polarized reflectivity setup scheme, (b) characterized fundamental mode resonance at telecommunications wavelength, blue dots represent data points, red line shows fit to a Fano-resonance with quality factor of $Q \sim 800$, (c) resonance intensity dependence on polarization angle of incoming light, blue dots represent data points, red line shows fit to $\sin(4\theta)$, where $\theta$ is half-wave plate rotation angle.

The observed resonances are blue-shifted relative to the theoretically predicted value by $\sim 90$ nm, while the Q factor is decreased by a factor of $\sim 1.75$. To explain these differences, we analyze the influences of material and fabrication imperfections, by simulating the resonant wavelengths and quality factors of photonic crystal cavities with parameters in the same range as those fabricated on the chip. We analyzed up to 30% hole radius increase, which could be caused in fabrication by over-etching, and the same percentage of slab thickness decrease, as the damaged interfacial layer leads to uncertainty in device layer thickness. The results are shown in Fig. 5 in green triangles and yellow squares, respectively. Both of these effects reduce quality factors and blue-shift the resonances, which is consistent with our experimental observations. Additional deviations may be caused by factors such as imperfect sidewall profile.[37]

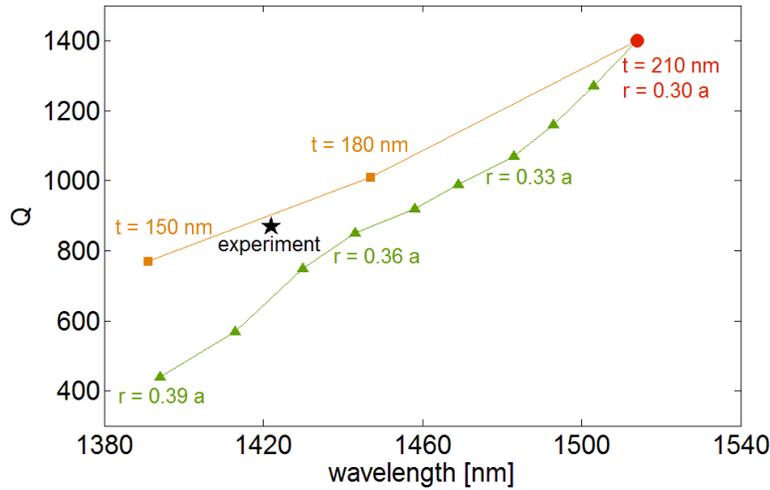

Fig. 5. Theoretical analysis of the wavelemgths and the quality factors of SiC L3 photonic crystal cavities as a function of the hole radius ($r$) shown in green triangles and slab thickness ($t$) shown in yellow squares. The optimal parameters ($t = 210$ nm, $r = 0.30\ a$) are represented by the red dot and the corresponding experimental result is indicated by the black star.

For most applications, it is necessary to be able to controllably target resonance wavelengths of high-Q cavities both over a large range and to specific wavelengths of interest (e.g. lines of optical emitters). In order to generate cavities that are tunable across the telecommunications band, we therefore varied the lattice constant $a$ of the L3 cavities in the range 500 – 700 nm. We again monitored the resonant wavelength and quality factor of the fundamental modes of these structures via the cross-polarized reflectivity technique. The results of this experiment, which are presented in Fig. 6, show that resonances of L3 cavities of Q ~ $10^3$ may be achieved across wavelengths 1,250 – 1,600 nm.

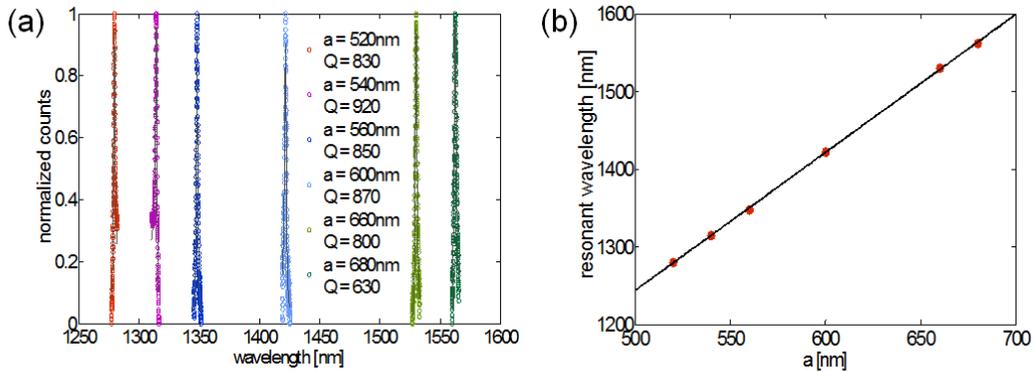

Fig. 6. (a) Fundamental mode resonance scaling with lattice constant $a$, shown with Fano-resonance fit with quality factors in range 630 to 920; all cavities are shifted L3 cavities made in 210 nm thick slab of 3C-SiC. (b) Experimentally measured linear dependence of resonant wavelength on lattice constant. Red circles are measured data points and black line is a linear fit to the data.

## 4. Conclusion

We have demonstrated a robust fabrication routine that has allowed us to realize a diverse set of devices such as 2D planar photonic crystal cavities, 1D nanobeam photonic crystal cavities, and crossbeam photonic crystal cavities from cubic (3C) silicon carbide films. We have also shown that the procedure may be used to realize high quality factor (Q ~ $10^3$) resonances at telecommunications wavelengths. In the future, we may overcome the low refractive index of 3C silicon carbide and offer ultrahigh-Q cavity design (Q ~ $10^6$) based on the nanobeam geometry.[38-40] The demonstrated scalable fabrication process will enable wider use of 3C-SiC nanophotonic classical and quantum information processing.

**Acknowledgements**

This work was supported by the Presidential Early Career Award for Scientists and Engineers (PECASE) from the Office of Naval Research and National Science Foundation (Grant ECCS-1025811), M.R. and S.B. were also supported by the Stanford Graduate Fellowships and National Science Graduate Fellowship. TMB was supported by the Nanoscale and Quantum Science and Engineering Postdoctoral Fellowship. We thank Sang-Yun Lee, Helmut Fedder, Jörg Wrachtrup, and Phillip Hemmer for helpful discussions. This work was performed in part at the Stanford Nanofabrication Facility of NNIN supported by the National Science Foundation under Grant No. ECS-9731293.